\documentclass[10pt]{revtex4}
\usepackage{ulem}
\usepackage{url}
\usepackage{graphics}

\begin{document}
\title{A civil super-Manhattan project in nuclear research for a safer and prosperous world}

\author{D. Sornette}
\affiliation{ETH Zurich\\Department of Management, Technology and Economics\\
Scheuchzerstrasse 7, CH-8092 Zurich, Switzerland\\
dsornette@ethz.ch}

\date{\today}

\begin{abstract}
Humankind is confronted with a ``nuclear stewardship curse'', facing the prospect of needing to
manage nuclear products over long time scales in the face of the short-time scales of human polities. 
I propose a super Manhattan-type effort 
to rejuvenate the nuclear energy industry to overcome the current dead-end in 
which it finds itself, and by force, humankind has trapped itself in. A 1\% GDP investment
over a decade in the main nuclear countries could boost 
economic growth with a focus on the real world, epitomised by nuclear physics/chemistry/engineering/economics
with well defined targets. By investing vigorously to obtain scientific and technological breakthroughs, 
we can create the spring of a world economic rebound based on new ways of 
exploiting nuclear energy, both more safely and more durably.
\end{abstract}

\maketitle

 
  \vskip 0.1cm
  \noindent
{\small I acknowledge useful feedbacks from Peter Cauwels, Tatyana Kovalenko, Wolfgang Kr\"oger, Paul-Emmanuel Sornette, Benjamin Sovacool, Hideki Takayasu and Spencer Wheatley.}

 
\section{Nature of the problem: societal risk and nuclear energy hazard}

Human development of nuclear materials for civil (as well as military) uses
has created a singular situation. Here the term ``singular'' is taken in the strong sense
of a unique situation with no equivalent, ever. The singular situation is that humankind
has put on herself the task of husbandry of nuclear materials and of the waste of civil and military uses
for centuries, tens of millennia and up to millions of years, depending on the nature
of the radioactive elements. Indeed, by-products of  a reactor last for hundred years 
 (e.g. Cesium-137 with a half-life of 30 years) to hundreds of thousand of years or even millions of years (Plutonium-239 with
 a half life of 24'000 years to Technetium-99 with the largest fission product yield of 6\% for 
 thermal neutron fission of Uranium-235 among long-lived fission products with a half life of 211'000 years). 

 \vskip 0.2cm
Consider that these time scales, during which humankind needs to babysit
these nuclear residues, are comparable to, or larger than, that of the lifetime of the human species `homo sapiens', 
usually dated to have emerged in his modern anatomical form about two hundred thousands years ago!
It is essential for the biosphere, and in particular for human health,
that the artificially concentrated and man-made nuclear materials are not entering the biological cycles.
The singularity of centuries, tens of millennia to the million-year time scales of required human management
stands in stark contrast with all other activities involving time scales of decades
to centuries -- at most, even in the worst chemical pollution cases. Even the long time scales
involved in global climate change are dwarfed by those resulting from human nuclear activities.

\vskip 0.2cm
It is thus essential to frame the issue of nuclear energy within the dynamic context of society.
Human societies are in continuous evolution, formation, aggregation, fusion, 
consolidation, disaggregation, collapses and so on. Human societies are punctuated by
transitions taking the form of revolutions, civil wars, conflicts, ethnic collisions and
instabilities (Cederman et al., 2013).  The typical time scales of human societies are decades to centuries, at best
(Turchin, 2003; Diamond, 2011; Fukuyama, 2014).
No empire, nation or society has ever been stable and without major conflicts over 
time scales of more than a few decades. 
Even the most stable society evolves. And a large body of evidence shows that these
evolutions often occur abruptly rather than through smooth transitions (Scheffer, 2009).
In other words, a large body of works on comparative history (Diamond and Robinson, 2011)
and, more recently, on geopolitical 
dynamics, show that human societies are not stable but are prone to crises and sudden mutations.

\vskip 0.2cm
Think of the crisis in Europe in the early 1990s, leading to the breakup of Yugoslavia following a series of political upheavals 
and inter-ethnic Yugoslav wars affecting primarily Bosnia and Croatia.
Consider the so-called Arab Spring started in December 2010, that destabilised regimes that 
had been stable for decades -- such as Tunisia, Egypt, Libya, Mali, Syria and Yemen -- and with major protests 
in a host of other Arab countries. Reflect on the situation in Pakistan, Bangladesh, Iran, Thailand, 
in many African countries and in South America; the political and social world is very far from stable,
even on a time scale of years to decades. And what about the growing inequalities
in the Western world in the last decades characterised by a progressive but steady impoverishment 
of the bottom 99\% of the population (Piketty, 2014), which may breed again new large scale instabilities with uncertain and
perhaps extraordinary consequences?
  
 \vskip 0.2cm
 
Humankind is thus faced with the problem of managing sensitive man-created wastes over much longer
time scales than the lifetime of ephemeral human polities. In fact, the development
of nuclear energy has been based on an implicit formidable bet that human societies will be sufficiently
stable, solid, and reliable to put nuclear husbandry at a suitable priority level, and at all times,
in order to avoid catastrophic singular events or a progressive alienation of our biosphere.
Nuclear energy is a recent phenomenon, whose development covers no more than the last
60 years or so. In this time, the major powers have been on the brink of 
total mutual destruction during the Cold War.  Many argue that the military nuclear threat
has been the very engine of stability in the Western world, 
following two debilitating (except for the US) world wars. But what about the more than 140 countries
on this planet that are
deemed non-democratic and exhibit various levels of potential or rampant instabilities?
And, when examined from a historical perspective, it may be no less than a heroic 
claim that societies that are now seen as stable will not transform into locii of
instability. 

 \vskip 0.2cm
According to another narrative, the end of WWII, followed by the bipolar world order organised around the two superpowers
resulting in the Cold War, led to an illusion of stability -- a dream that social and political systems
have evolved towards higher levels that could ensure better outcomes to resolve
human conflicts. However, History suggests that betting on human peace and stability may be dangerous.
Perhaps, the situation is becoming even more uncertain, with the progressive 
transition to a new regime where scarcity of natural resources and essential
commodities and the competition for vital space will shape the new densely populated world order.

 \vskip 0.2cm
How is it possible to ensure that teams of skilled technicians will dutifully 
continue their routine maintenance of key nuclear facilities and waste
storage sites in the presence of a local revolution, conflict or war threatening their families?
What if Saddam Hussein, exasperated after losing power, had 
a nuclear power plant (NPP) to make critical (by simply destroying or incapacitating the cooling systems), 
rather than burning oil fields?
Even worse, in the event of severe conflicts between nations, NPP 
and other critical infrastructures become prime targets
in the goal of crippling the enemy.  As a recent vivid illustration, during the Ukrainian civil war,
there was active social media activity concerning the calls to attack 
the Zaropozhskay NPP (the largest NPP in Europe and the fifth largest in the world),
which is 200 km from the war zone.
In February 2014, operatives of the Right Sector were arrested by guards of NPP when trying to infiltrate them,
forcing NATO nuclear specialists to check that all Ukrainian NPPs have adequate protection measures.

\vskip 0.2cm

Another dimension of the singularity of nuclear energy is the extraordinarily large impact
that a single accident can have at the worldwide level. 
There are currently more than $440$ nuclear reactors in operation and more than $60$ 
under construction worldwide. For all, one cannot exclude the possibility of another accident involving a partial 
meltdown of the reactor of a large NPP, with a significant fraction (say 5\% to 20\%) of the reactor 
contaminating the atmosphere, ocean and/or Earth soil. Our estimates show that just one event 
has global measurable consequences (Sornette et al., 2013; Wheatley et al., 2015). 
Supposedly impossible scenarios (according to industry-standard Probabilistic Safety Analysis risk estimates)
such as Chernobyl and Fukushima (Kr\"oger and Sornette, 2013)  can be taken as the basis to imagine others, whose
impact would be in the range of tens of trillions (of dollars, euros, Swiss francs...) with 
lasting consequences in the form of major zones of uninhabitability (Sornette et al., 2013; Wheatley et al., 2015). 
Think for instance of the real-estate value of 
New York City, USA or of Zurich, Switzerland, both of which are rather close to an operating NPP
and would become uninhabitable in exceptional and extremely unlikely  -- but possible -- scenarios.  
Hence, the management of the nuclear energy industry should be considered as a public good, 
where any accident or misbehavior in one major NPP has externalities over the whole planet. 
Most relevant to Europe, the Chernobyl accident had -- and still has -- significant environmental, 
health and financial implications. Furthermore, there is no guarantee 
that radioactive materials will remain hermetically enclosed in the concrete sarcophagus in the future. 
This is a Damocles sword hanging over the head of large European populations for many generations.  
Thus, even in stable society, nuclear risk is still high (Sornette et al., 2013; Wheatley et al., 2015).
The official industry and political position is a combination of dangerously underestimating risk, 
and being disingenuous about it to the public.  Clearly the risk needs to be understood and publicly 
acknowledged before there will be public support for addressing it.

 \vskip 0.2cm
Even in supposedly stable and efficient societies, it is doubtful that we can count 
on the reliability of human managed organisations to ensure a safe nuclear stewardship.
In his study of the safety of the US nuclear weapons command organisations, Scott Sagan (1995)
 provided numerous examples that the organisations one may have surmised 
to be those with the best safety record are in fact plagued by failures and accidents,
due to political infighting, organized deception, normalization of errors, reclassification of 
failure as success, and conflicts over short-term interests. 
In a recent book, Eric Schlosser (2014) goes further by reporting in details on known 
accidents with nuclear weapons that have been regularly taking place since 1945. 
Centering on the Damascus accident of 1980, an explosion in a Titan II Inter-Continental Ballistic 
Missile housed in Damascus, Schlosser documents a litany of nuclear accidents revealing 
the past, present and future vulnerability of the exceedingly complicated technical 
systems that are nuclear weapons, embedded within layers of bureaucracy and subjected 
to the continuously changing nuclear policies of the politicians. The civil nuclear energy
industry is distinct from the nuclear weapon organisations, but for our purpose they both share 
a number of important characteristics:
(i) they deal with the same high energy density of nuclear physics at 1 MeV; (ii) they deal with extremely complex
systems, interfaced with and managed by fallible human operators embedded in imperfect institutions; (iii) they are exposed to the changing whims of politicians, themselves
reacting to the volatile public opinions. 

\vskip 0.2cm
And there is the controversial economics of nuclear energy
(see e.g. \url{http://www.world-nuclear.org/info/Economic-Aspects/Economics-of-Nuclear-Power}
and Shrader-Frechette, 2011). On March 4, 2015, the french Areva nuclear group announced
record losses of 4.8 billion euro for 2014. One may wonder whether and when cost-cutting will impact 
security measures and reduction of competent personnel?  
Most of the $440$ NPP in operation are over 30 years old and will require increasing investment
to ensure safety, not to speak of the cost of decommissioning a NPP,
which may turn out to be roughly on par with the cost of building it.

\vskip 0.2cm
 In the best case scenario (in the absence of conflict and regime change),
 according to a detailed statistical study of the most complete available database that is 75 percent larger than the previous best dataset on nuclear incidents and accidents, Wheatley et al. (2015) found that we still have a a 50\% chance that, in terms of costs, 
 (i) a Fukushima event (or larger) occurs in the next 50 years, (ii) a Chernobyl event (or larger) occurs 
 in the next 27 years and (iii) a TMI event (or larger) occurs in the next 12 years. The figure provides
 a precise statistical quantification of these statements.
 This suggests an intrinsic instability of the nuclear energy industry. Together with my other argument
on the instability of societies, the diagnostic is inescapable: an unstable industry in an unstable world.

\begin{figure}
	\includegraphics{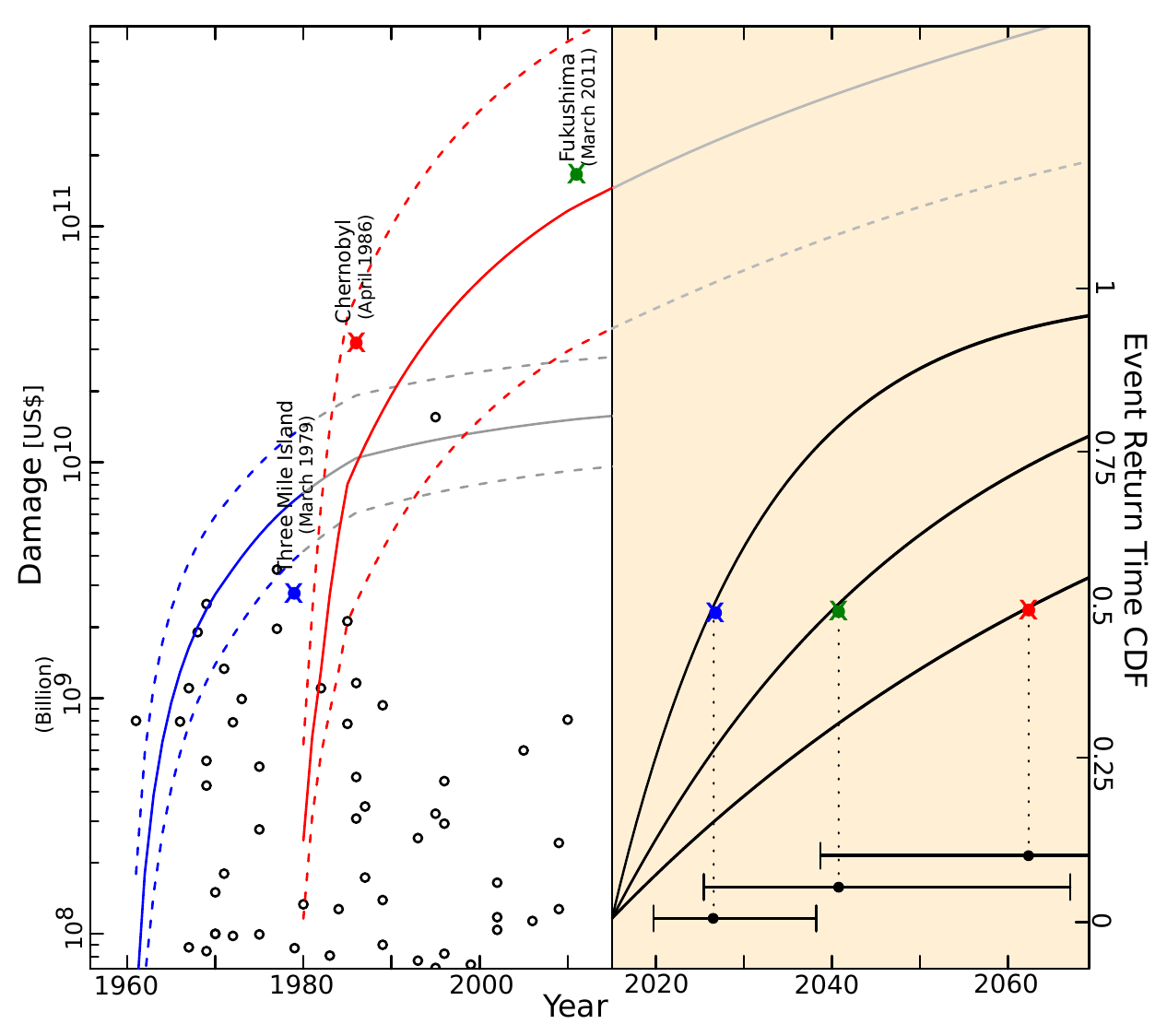}
	\caption{{\bf Left panel}: Left panel: Historic data (open circles) showing the US\$ losses (left vertical scale) resulting from accidents, in excess of 70 million US\$, occurring at nuclear power generating facilities,
 between 1960 and 2014. The data is compared with the median and quartiles of the estimated distribution of the largest extreme events, for the pre- (blue) and post- (red) TMI (Three Mile Island accident of 1979) regimes, using methods described in (Wheatley et al., 2015). Had the distribution not changed after TMI, we would expect the largest events (grey) to
be much smaller than under the post-TMI regime. 
{\bf Right panel}: cumulative distribution function (right vertical scale) of the event arrival time for a Fukushima like (or larger US\$ cost) event, a Chernobyl like (or larger US\$ cost) event, and a TMI like (or larger US\$ cost) event. The 50\% arrival times are indicated by the x. Beneath, the 25\%-75\% arrival interval is given for each case. This visualises the statement in the text and further quantifies the 25\%, median, and 75\% quantile return times: Fukushima: 2015 + (25, 50, 225); Chernobyl: 2015 + (11, 27, 55); TMI: 2015 + (5, 12, 25). Figure prepared by S. Wheatley.}
	\label{fig_MaxGrowthReturn}
\end{figure}
 
 \vskip 0.2cm
I then ask how can the reliable management of nuclear risks
be achieved over the required time scales of tens of years, hundreds of years, 
up to millions of years, given the fundamental unstable
nature of human societies and polities (Sornette and Cauwels, 2015).
I contend that these simple facts (long time scales of needed nuclear stewardship,
short-time scales of human polities and fallibility of technical human organisations) 
when put together, lead to a deep reassessment 
of humankind's choices with respect to the nuclear industry.


\section{Proposed solution and action: call for a super Manhattan-type effort in nuclear research and
engineering}

This communication is not a call to stop the development and use of nuclear energy.
In view of the diagnostics of the previous section of our stewardship curse, I conclude that 
there is no way to turn back, given the existing millions 
of tons of radioactive substances
(with a complex hierarchy of radioactivity levels and lifetimes) that have already been created
by the half-century operation of hundreds of NPP (in addition to military grade substances).
It is not like we could vote ourselves out of the problem by choosing to get out of nuclear energy,
as naively proposed by some green advocates, policy makers and governments.
This is simply impossible as the unnaturally concentrated nuclear material (in wastes and NPP) is with us for up to
millions of years. Phasing out nuclear energy will not make the nuclear wastes, and
the radioactive elements associated with the decommissioning of the NPP, disappear.

\vskip 0.2cm

Given the highly unstable nature of human societies, I conclude that it is simply irresponsible to depend
on the reliability of human management of these delicate infrastructures and wastes
in the long and very long terms. What are the alternatives? My proposal is straightforward and
probably will be called naive and unrealistic by many: let us get
rid of the wastes, of the past, present and as it is generated. This is the only logical way
to address the dual problem of the unreliability of human management and the presence of
critically dangerous long-lived wastes.  This requires directly attacking 
the key social problems confronting nuclear power, which are poor economics, the possibility of 
catastrophic accidents, radioactive waste production, and linkage to nuclear weapon proliferation
(Ramana and Mian, 2014). By making clean civil nuclear energy a priority, we ensure
the necessary development and sustainability of the technological and human skills.
In contrast, phasing out civil nuclear energy leads to the enormous risk of losing 
the critical human competence to ensure the long-term stewardship of the 
mountains of nuclear wastes, in a society that would inherit a costly legacy 
of previous mistaken policies and would no more be committed to this energy source. Nuclear energy is a fact of Nature.
We can ignore it at the cost of a huge opportunity loss. We can mistreat it
at the cost of enormous dangers. But it is there and we have not done enough yet to master it to 
garner the fruits of this amazing source of energy. Now is the time.

\vskip 0.2cm

Achieving this goal is neither trivial, nor is success certain. But not trying would be failing
to embrace the option that can free us from the nuclear stewardship curse
we put ourselves in. It requires a vigorous civil nuclear research program
to crack the waste recycling problems. I propose that 
governments should  immediately invest large resources in nuclear physics,  
nuclear engineering and chemistry (1 MeV energy scale),
relevant ordinary chemistry (1 eV energy scale)
as well as material sciences and relevant engineering disciplines  to find innovative ways of solving
the key problem of the long-lived radionuclides. I am in particular thinking of both incremental progress
as well as the development of completely novel revolutionary methods for 
energy-efficient transmutation with accelerators and other novel technologies. 
Recall that, with large flux of neutrons, it is in principle possible to reduce the half-life of Plutonium
from 24'000 years to just one day. 
Accelerator driven transmutation systems have been proposed already a long time ago (see e.g.
Bowman et al., 1992)
and the impact of this and other advanced waste management strategies 
and closed fuel cycle partitioning and transmutation technologies on the long term radio-toxicity of 
stored waste have already been the subject of numerous investigations
(see e.g. Saito and Sawada, 2002; OECD Nuclear Energy Agency, 2002; 
OECD NEA No. 6996, 2012; Knebel et al., 2013; Abderrahim et al., 2014).  We need
to build on this existing knowledge and leverage it many times to foster major
innovations necessary to make this practically feasible 
and economically viable, even profitable. Further, there is a major political obstacle, which is the need for
the prevention of nuclear weapon proliferation, which was at the origin of US President Carter's
decision on April 27, 1977 to indefinitely defer the reprocessing of spent nuclear reactor fuel 
by so-called breeder reactors (see (Rossin, 1995) and President Carter's message to Congress
entitled  ``The need to halt nuclear proliferation is one of humankind's most pressing challenges'').
Thus, with concern to the full nuclear fuel cycle, it will be essential to 
work on reprocessing, breeder reactors, nonproliferation and Plutonium honestly,  openly and objectively. 
 
\vskip 0.2cm
Due to the need to recycle piles of nuclear waste, we need more than just incremental improvements in 
Generation IV and V reactors to achieve energy production that is more economical, safer, 
minimizes ard reduces waste, and avoids proliferation. This is because
of the need to recycle the piles of nuclear wastes and get rid of the stewardship curse
that we are currently in. Our present mastering of nuclear energy
is still quite primitive and wasteful. We need to develop the technologies for efficient reliable
transmutation using systems that work in sub-critical and/or critical mode, in the spirit 
of the intended European MYRRHA project, but many times more ambitious and faster.
The tasks are manifold, involving reactor technology for safety and efficiency, 
reprocessing physics and engineering, new materials for thermal exchanges, 
techniques of energy production, risk mapping, and so on.
This will require a strong multidisciplinary and interdisciplinary approach that brings closely together
nuclear physicists, engineers, chemists, natural and Earth scientists, economists, 
as well a political scientists, conflict research scientists and historians.
The quasi-abandoned nuclear physics and chemistry research fields
should be rekindled  in  major universities worldwide. We need a renaissance of nuclear science, 
with a resurrection of  moribund nuclear physics departments in higher education. 
The private sector has to be incentivised even further to contribute its share 
of funding and creativity in the goal of using the new techno-innovations for efficient and 
reliable nuclear energy sources. 

\vskip 0.2cm
A major quantitative map must be developed, combining the knowledge
on nuclear risks from an engineering point of view, also taking into account the 
impact of natural hazards (earthquakes, tsunamis, storms, floods...) and geopolitical and 
economic risks. On top of this engineering map, a major interdisciplinary effort is needed to identify
all kinds of social instabilities. A lot of data and knowledge already exist along
these different dimensions  but it needs
consolidation, integration and enhancement in order to become operational.
By combining these different elements, we shall
obtain the first comprehensive risk map associated with nuclear materials created by humankind
to better manage in an integrative fashion the existing installations and better develop future ones.

\vskip 0.2cm
There are a number of existing activities exploring options for transmutations in Europe and 
worldwide (Struwe and Somers, 2008). These initiatives demonstrate
the existence of high uncertainty with respect to a host of 
physical and technical problems. Making nuclear energy clean and reliable is 
an inherently complex system problem, given the multiple scales, interactions and feedback loops involved, and
including the notorious human-machine interfaces and the tendency for humans to make mistakes.
Broad R\&D programs are needed to  invent, explore and
evaluate open material properties for the reliable operation of these complex systems.
The challenges are not just technical but also theoretical, with the need for the
development of multidimensional theoretical approaches to 
improve the understanding of fuel behaviour and reactor processes and to foster
novel reactor designs for optimal transmutation. Alternative technologies such as 
thorium-fuelled molten salt reactors need to be investigated and improved, since thorium is about three
times as abundant as uranium, does not lead to core melt down and produces less waste,
with the present designs (Tindale, 2011). Approaches such as the 
Electron Machine with Many Applications (EMMA) could lead to accelerator driven subcritical reactor systems
in which a non critical fission core is driven to criticality by a small accelerator (Rose, 2011)
(see also \url{http://en.wikipedia.org/wiki/EMMA_%28accelerator%29}). 
In the last few years, the nuclear industry has been advocating the `new' technology of 
small modular reactors, on the claim that it could overcome previous economic failures of nuclear power. 
However, recent studies suggest that enormous efforts are required to reverse the severe
disadvantages that small modular reactor technology may face in the future 
(Ramana and Mian, 2014; Cooper, 2014). This field therefore needs more vigorous research.


\section{Broader economic context and vision}

As explained in (Sornette and Cauwels, 2014), the period from 1980 to
the great financial crisis starting in 2008 was characterised by growth
that was mainly catalysed by excessive unsustainable private sector debt, financial
innovation and leverage. Since 2008, and especially in Europe, one can observe
sluggish growth or even stagnant economies, with a strong concern that 
the growth that the West has enjoyed since the reconstruction era
after WWII might be a phenomenon of the past, given the levelling off
of demographics and the slowing down of technological innovations (Gordon, 2012).
Given that the post-crisis solutions have been even more debts on the side
of governments and easy monetary policy -- the very ingredients that facilitated
the run-up to the crisis --, how can we get out of this ``new normal'', and face the
enormous liabilities of our underfunded pension systems?

\vskip 0.2cm

I propose a bold initiative for governments and society to invest confidently in the future
with a $\sim 2\%$ additional GDP investment focused on energy, from nuclear energy to 
diluted sources such as solar, wind, geothermal, as well as the battery storage problem and so on. 
Humankind's achievements have
been decided by energy. Most of human history has been determined by human and animal energy.
Then, came the industrial revolution of the eighteen century with a transition to coal energy
and then to oil energy (Yergin, 1991), and to other sources. It is often remarked
that a typical European or American household possesses more artefacts
than an Egyptian pharaoh, simply by its access to sources of high density energy.
We tend to take it for granted, but our civilisation is one of dense and plentiful energy.
We thus need to foster innovation in energy of all forms, both dense and more centralised
like nuclear energy that complement the more dilute and decentralised forms.
As described above, I view the nuclear energy challenge as preponderant, which should garner a dominant
share, perhaps as much as 1\% of GDP for a decade and more.

\vskip 0.2cm
Rather than bailing out large existing banks and giant dinosaur-like firms, we should be wiser to foster
the development of small and medium-sized firms, which are well-known to be
the real creators of jobs (Wong et al., 2005; van Praag, 2007) (while large firms are net destructors of jobs, as a result
of consolidation during M\&A and cost cutting). While most of civil nuclear energy
has tended to be centralised in large firms and government agencies, there is a place
for bottom-up innovations and bold initiatives, in combination with carefully thought
industry-public partnerships. This would send a strong
message to the tens of millions of young people in Europe and in the US in particular
that they are encouraged and helped to innovate, to take risks and to join the
dream of the previous generation of creating a better future.  
This strategy would help reverse the negative sentiments that have
been permeating industry and political circles, as well as the heavy 
economic and psychological load carried by households, via the 
message of hope and the vision of a better future.
This requires explaining the importance of expectations, of self-fulfilling anticipations
and of providing the incentives to create a new real economy. 
There is empirical evidence
that the successful future companies are often those created during recessions 
and during bad times by entrepreneurs who are contrarians and dauntless.
The investments proposed by this program could help support such a recovery.

\vskip 0.2cm

The exceptional economic and political situation in which 
the world, and Europe in particular, finds itself, calls for exceptional measures and ideas.
In the end, there is only one way to get out of debt and over leverage, which is growth.
Countries have rarely if ever paid off their debt, they have grown their GDP out of their debt.
I mean real growth, and not the artificial unsustainable financial based leverage that led to the 2008 banking crisis
in the US and the 2010 sovereign troubles in Europe that are still continuing and far from 
being resolved (Sornette and Cauwels, 2014). 
There are only two ways to grow (per capita) in the real economy (as opposed to the financial virtual economy of the 
past three decades before 2008), which is to create/innovate and increase productivity.
This can only be achieved by humans for humans, in the presence of the appropriate infrastructure
and the right vision on great goals.
This will require special measures at the fiscal and credit levels to facilitate the drive of 
the energetic part of the population, to raise its head again, take the bull by the horns,
and innovate and create.  
More broadly, this vision weaves with the construction of
a future of new energies, of new transportation and network infrastructures, of new cities and dwellings
(better integrated with natural habitats), of new health and food  industries, 
of new entertainment, and so on. Special attention and processes must be put in place
to include the public wellbeing at the center of the initiative, to empower people so that they 
can develop trust in the specialised professions and organisations dealing with nuclear
energy for their competence and communication honesty (Greenberg et al., 2014),
which has, in many cases, arguably been badly managed: for instance, 
the International Atomic Energy Agency (IAEA) does not publish a historical database of the events
rated on the International Nuclear Event Scale, leading to a gap in the public perception between the claimed extremely low
likelihood of a core melting accident and the occurrence of several of them already in just a few
decades of nuclear history.  Only then
can the required effort be accepted as a policy priority (Greenberg, 2014), not top down but bottom up,
from the people, by the people, for the people.  More generally, progress in public infrastructure would help
reverse the disenfranchisement that people are feeling due to government mal investment.

\vskip 0.2cm

As already mentioned, the investment in a vigorous civil nuclear research program should be the priority.
I indicated a rough figure of about 1\% of GDP per year for a decade.
How does this compare with past projects?
To give a first reference, the Manhattan project cost a total \$2 Billion (which should be compared with 
the US 1940 GDP of \$100 Billlion and 1944 GDP of 220 Billions that rose tremendously 
just in four years due to  the WWII effort), amounting to about US1998 \$26 Billion (Schwartz, 1998) (compared
to a 1998 GDP of \$9'000 Billions). The Manhattan project was thus 2\% of the US 1940 GDP and
1\% of US 1944 GDP (but of course spread over several years). 
The Apollo program cost \$30 billion in 1970s dollars,
the equivalent of \$110 billion in 2010. The total cost of the space senhuttle program was estimated
at close to \$200 billion in 2010 dollars (\url{http://www.thespacereview.com/article/1579/1}). 
The US has spent \$486 billion over 57 years on human spaceflight, an average of \$8.5 billion a year.
A recent report prepared by the US Congressional Research Service  (Sissine, 2014)
quantified that the funding in nuclear energy technology by the US Department
of Energy (DOE) summed up to 50 billion of 2013 US\$ for the 37 year period from FY1978 to FY2014.

\vskip 0.2cm
Compared with this paltry 1.35 billion of 2013 US\$ per year of spending of the US DOE, I propose a hundredfold increase
in each of the major nuclear energy countries (USA, Europe, Japan, China...).
One hundred fifty billion US\$ per year seems huge, but is a tiny fraction of the bailout program launched in 2008
involving trillions of dollars to save financial institutions during and after the financial crisis. 
It is also slightly less than 1\% of US 2014 GDP.
One hundred and fifty billion euros
is more than half the present GDP of Greece, but it amounts to just  two months and a half
of the ECB quantitative Easing program launched in March 2015 on the tune
of 60 billions euros per month to buy European sovereign bonds.
Eighteen trillion Yen seem astronomical but is dwarfed by the expansion of the Bank of Japan
bond buying program at a rate of 80 trillion Yen of bonds a year, as of 31 October 2014
(\url{https://www.boj.or.jp/en/announcements/release_2014/k141031a.pdf}). 

\vskip 0.2cm

The picture is clear: to keep the status quo of the financial and economic states
of their countries, present high level officials do not hesitate to
develop completely non-standard policies, whose real economic impacts are highly controversial, 
promoting moral hazard, catalysing misallocations of resources, launching currency wars, increasing
inequalities and leading to
a full range of unintended consequences and unknowns. These so-called ``quantitative easing''
programs use the levers of finance and innovative monetary policy in the hope that this
will spill-over to the real economy and spur job recovery and real economic growth.
In fact, such quantitative easing worsens inequality, thus making social instability more likely,
exacerbating the precariousness of civil nuclear energy.
In contrast, my proposition targets the real world directly, epitomised by nuclear physics/chemistry/engineering
with well defined targets. By investing vigorously to obtain scientific and technological breakthroughs, 
we create the spring of a world economic rebound based on new ways of 
exploiting nuclear energy as well as all other possible energy sources, both more safely and more durably
(Kovalenko and Sornette, 2013).

\vskip 0.2cm

A big part of my proposed program includes a new system of incentives to attract
the best minds to the civil nuclear research and industry renaissance.
It has been well publicised that the past 50 years have seen an almost doubling of the 
income of people employed in the banking and financial industry compared
with people employed in other industries, with the same level and sophistication of education.
As a professor of finance (but also of Physics, of Earth Sciences, of Complex Systems
and of Entrepreneurial Risks) at the Swiss Federal Institute of Technology in Zurich (ETH Zurich), I witness
daily the dramatic attraction that finance has on some of my best students, lured
by the great technical challenges, high degree of innovation, glittering jobs and high incomes.
While financial innovation is necessary to solve funding problems for the development
of new technologies (Janeway, 2012), it is not the solution for our real concrete problems
and hopes in the real economy, namely sustainability with respect to water, food, energy,
access to new horizons
of entertainments and technology, and the pursuit of happiness.
Thus, we need strong policies to reverse the wrong incentives that attract the best
minds to Wall Street and to redirect them to the real challenges, in particular of
nuclear science and technology. This requires a revamping of the salary and funding system, starting with
departments of top universities to the development of proper stimuli in the industry.
Let us attack courageously the misaligned incentives that have created so many
moral hazards and misallocation of resources and talents to focus on the great challenges of our time.

\vskip 1cm
\noindent
{\bf References}

Abderrahim, H.A., D. De Bruyn, G. Van den Eynde and S. Michiels,
Transmutation of high-level nuclear waste by means of accelerator driven system,
WIREs Energy Environ 3, 60- 69 (2014).

Bowman, C.D., E.D. Arthur, P.W. Lisowski, G.P. Lawrence, R.J. Jensen, J .L. Anderson,
B. Blind, M. Cappiello, J.W. Davidson, T.R. England, L.N. Engel, R.C. Haight,
H.G. Hughes III, J.R. Ireland, R.A. Krakowski, R.J . LaBauve, B.C. Letellier, R.T. Perry,
G.J. Russell, K.P. Staudhammer, G. Versamis and W.B. Wilson,
Nuclear energy generation and waste transmuation using an accelerator-driven intense thermal
neutron source, Nuclear Instruments and Methods in Physics Research A 320, 336-367 (1992).

Cederman, L.-E., K.S. Gleditsch and H. Buhaug, Inequality, Grievances, and Civil War 
(Cambridge Studies in Contentious Politics) Cambridge University Press (August 26, 2013).

Cooper, M., Small modular reactors and the future of nuclear power in the United States,
Energy Research \& Social Science 3, 161-177 (2014).

Diamond, J., Collapse: How Societies Choose to Fail or Succeed: Revised Edition,
Penguin Books; Revised edition (January 4, 2011).

Diamond, J. and J.A. Robinson, eds., Natural Experiments of History, Belknap Press (2011).
 
Fukuyama, F.,  Political Order and Political Decay: From the Industrial Revolution to the Globalization of Democracy,
Farrar, Straus and Giroux (September 30, 2014).

Gordon, R., Why Innovation Won't Save Us, The Wall Street Journal, Dec. 21, 2012.

Greenberg, M.R., Energy policy and research: The underappreciation of trust,
Energy Research \& Social Science 1, 152-160 (2014).

Greenberg, M.R., M.D. Weiner, D. Kosson and C. Powers,
Trust in the U.S. Department of Energy: A post-Fukushima rebound,
Energy Research \& Social Science 2, 145-147 (2014).

Janeway, W.H., Doing Capitalism in the Innovation Economy,
Cambridge University Press (October 8, 2012).

Knebel, J., C. Fazio, W. Tromm and W. Maschek,  What to do with radioactive waste?
Spektrum der Wissenschaft 44 (11), 34-39 (2013) (in german)

Kovalenko, T. and D. Sornette, Dynamical Diagnosis and Solutions for Resilient Natural and Social Systems,
Planet\@ Risk 1 (1), 7-33 (2013) Davos, Global Risk Forum (GRF) Davos
(http://arxiv.org/abs/1211.1949)

Kr\"oger, K. and D. Sornette, Reflections on Limitations of Current PSA Methodology,
ANS PSA 2013 International Topical Meeting on Probabilistic Safety Assessment and Analysis,
Columbia, South Carolina, USA, September 22-26, 2013, American Nuclear Society, LaGrange Park, IL (2013),
invited article for the Probabilistic Safety Analysis 2013 (PSA2013) (accepted 5 July 2013)

OECD Nuclear Energy Agency (2002), 
Accelerator-driven Systems (ADS) and Fast Reactors (FR) in Advanced Nuclear Fuel Cycles Ð A Comparative Study,  available on the NEA webpage on Accelerator-driven Systems (ADS) and Fast Reactors (FR) in Advanced Nuclear Fuel Cycles (\url{https://www.oecd-nea.org/ndd/reports/2002/nea3109-ads.pdf})
 
OECD NEA No. 6996, Actinide and Fission Product Partitioning and Transmutation,
Nuclear Science and Nuclear Development,
Eleventh Information Exchange Meeting, San Francisco, California, USA, 1-4 November 2010,
Nuclear Energy Agency,   (ISBN 978-92-64-99174-3)  (2012).

Piketty, T., Capital in the twenty-first century, The Belknap Press of Harvard University Press 
(Cambridge, Massachusetts and London, UK, 2014).

Ramana, M.V.  and Z. Mian, One size doesnÕt fit all: Social priorities and technical 
conflicts for small modular reactors, Energy Research \& Social Science 2, 115-124 (2014).

Rose, D., This is Emma. She's going to save the world (and cure cancer), Mail Online (12 June 2011)
(\url{http://www.dailymail.co.uk/home/moslive/article-2001548/Electron-Model-Many-Applications-Technology-save-world.html#ixzz3Wbk9idVb})
 
Rossin, D., U.S. policy on spent fuel reprocessing: the issues, Frontline, PBS (1995)
\url{http://www.pbs.org/wgbh/pages/frontline/shows/reaction/readings/rossin.html}

Sagan, S.D., The Limits of Safety, Princeton University Press; 1 edition (January 9, 1995).

Saito, M. and T. Sawada (editors),
Advanced Nuclear Energy Systems Toward Zero Release of Radioactive Wastes Hardcover,
Pergamon; 1 edition (November 25, 2002).

Scheffer, M., Critical Transitions in Nature and Society, (Princeton Studies in Complexity),
Princeton University Press (July 26, 2009).

Schlosser, E., Command and Control: Nuclear Weapons, the Damascus Accident, and the Illusion of Safety, Penguin Books; Reprint edition (August 26, 2014).

Schwartz, S.I., Atomic Audit: The Costs and Consequences of US Nuclear Weapons. Washington, D.C.: 
Brookings Institution Press (1998).

Shrader-Frechette, K., Climate Change, Nuclear Economics and Conflicts of Interest,
Science and Engineering Ethics 17 (1). doi:10.1007/s11948-009-9181-y  (March 2011)

Sissine, F. Renewable Energy R\&D Funding History: A Comparison with Funding for Nuclear
Energy, Fossil Energy, and Energy Efficiency R\&D, Specialist in Energy Policy,
Congressional Research Service, 7-5700 (www.crs.gov), RS22858 (October 10, 2014)

Sornette, D. and P. Cauwels,
1980-2008: The Illusion of the Perpetual Money Machine and what it bodes for the future,
Risks 2, 103-131 (2014)
(\url{http://arxiv.org/abs/1212.2833} and \url{http://ssrn.com/abstract=2191509})

Sornette, D. and P. Cauwels,
A Creepy World: How can managers spot and manage systemic crises,
Journal of Risk Management in Financial Institutions (JRMFI) 8 (1), (2015)
(\url{http://arxiv.org/abs/1401.3281} and \url{http://ssrn.com/abstract=2388739})

Sornette, D., T. Maillart and W. Kr\"oger, 
Exploring the limits of safety analysis in complex technological systems,
International Journal of Disaster Risk Reduction 6, 59-66 (2013).

Struwe, D. and Somers, J., Overview of Activities in Europe Exploring Options for Transmutation,
Euradwaste '08, Seventh European Commission Conference on the Management and Disposal of Radioactive Waste,
(October 2008)
\url{ftp://ftp.cordis.europa.eu/pub/fp7/fission/docs/euradwaste08/papers/paper-8-overview-of-activities-in-europe-d-struwe_en.pdf}
 
Tindale, S., Thorium: how to save Europe's nuclear revival, Centre for European Forum,
 Policy Brief, London, UK (2011) (\url{www.er.org.uk}).

Turchin, P., Historical Dynamics: Why States Rise and Fall. Princeton, NJ: Princeton University Press (2003).

van Praag, M., What is the Value of Entrepreneurship? A Review of Recent Research (2007)
\url{http://papers.ssrn.com/sol3/papers.cfm?abstract_id=1010568}

Wheatley, S., B. Sovacool and D. Sornette,
Of Disasters and Dragon Kings: A Statistical Analysis of Nuclear Power Incidents \& Accidents,
Risk Analysis (submitted 7 April 2015) (http://arxiv.org/abs/1504.02380)

Wong, P.K., Y.P. Ho and E. Autio, Entrepreneurship, Innovation and Economic Growth:
Evidence from GEM data, Small Business Economics 24, 335-350 (2005).

Yergin, D., The prize (the epic quest for oil, money \& power), Simon \& Schuster, New York (1991)

\end{document}